# Detectors for the Gamma-Ray Resonant Absorption (GRA) Method of Explosives Detection in Cargo: A Comparative Study

David Vartsky[a], Mark B. Goldberg*[a], Gideon Engler[a], Asher Shor[a], Aharon Goldschmidt[a],
Gennady Feldman[a], Doron Bar[a], Itzhak Orion[b] and Lucian Wielopolski[c]

[a]Soreq NRC, Yavne 81800, Israel
[b]Ben Gurion University, Beer Sheba 84105, Israel
[c]Brookhaven National Laboratory, Upton N.Y. 11973, USA

## ABSTRACT

Gamma-Ray Resonant Absorption (GRA) is an automatic-decision radiographic screening technique that combines high radiation penetration with very good sensitivity and specificity to nitrogenous explosives. The method is particularly well-suited to inspection of large, massive objects (since the resonant γ-ray probe is at 9.17 MeV) such as aviation and marine containers, heavy vehicles and railroad cars. Two kinds of γ-ray detectors have been employed to date in GRA systems: 1) Resonant-response nitrogen-rich liquid scintillators and 2) BGO detectors. This paper analyses and compares the response of these detector-types to the resonant radiation, in terms of single-pixel figures of merit. The latter are sensitive not only to detector response, but also to accelerator-beam quality, via the properties of the nuclear reaction that produces the resonant-γ-rays. Generally, resonant detectors give rise to much higher nitrogen-contrast sensitivity in the radiographic image than their non-resonant detector counterparts and furthermore, do not require proton beams of high energy-resolution. By comparison, the non-resonant detectors have higher γ-detection efficiency, but their contrast sensitivity is very sensitive to the quality of the accelerator beam. Implications of these detector/accelerator characteristics for eventual GRA field systems are discussed.

**Keywords:** γ-Ray Nuclear Resonance Absorption (GRA), explosives detection, resonant detectors, BGO detectors

## 1. INTRODUCTION

GRA is an element-specific, radiographic-transmission imaging technique, capable of revealing the spatial density distribution of the element in question within inspected objects. The method was proposed by Soreq NRC (SNRC) in 1985 to the U.S. Federal Aviation Administration (FAA) as a means of detecting explosives in airline baggage via nitrogen-specific radiographic imaging [1]. Nitrogen GRA holds promise for an automatic explosives detection system (EDS), since high nitrogen density is a characteristic of most explosives used by terrorists, but not of commonly-transported benign materials.

The FAA sponsored and funded this project starting from a basic feasibility study[2], launched in 1987 in collaboration with Los Alamos National Laboratory (LANL), up to the stage of constructing a partial field-prototype inspection system. The latter underwent an extensive blind capability test in 1993 using an RFQ accelerator (to generate the requisite resonant γ-rays via proton-capture on a $^{13}$C target at the $E_p$ =1.746 MeV resonance) at LANL, yielding highly encouraging results [3,4].

For this system SNRC developed and manufactured γ-ray detectors with resonant response, that are selectively sensitive to photons in the 9.17 MeV region, which corresponds to an excited state in $^{14}$N. These detectors are based on liquid scintillators that contain a nitrogen-rich cocktail [5]. In parallel, the LANL group developed its own GRA system based on commercial bismuth germanate (BGO) detectors. The LANL prototype was tested in parallel to the SNRC prototype in the above-mentioned 1993 test on the same RFQ accelerator [6]. In 1994, another group consisting of a collaboration between TRIUMF and Northrop Grumman began developing a GRA contraband detection system based on segmented position-sensitive BGO detectors and an electrostatic Tandem accelerator [7,8].

* goldberg@soreq.gov.il; phone (972)-8-943 4750; fax (972)-8-943 4676

GRA is particularly well-suited to inspection of large, massive objects (the resonant 9.17 MeV γ-ray probe being highly penetrating) such as aviation and marine containers, heavy vehicles and railroad cars. Thus, the SNRC group reoriented its efforts in 1994 from inspection of individual airline passenger bags to aviation cargo containers. With the above-mentioned resonant-response detectors, a prototype for inspecting an LD-3 container was built and tested in 1998, using a proton beam from the Dynamitron accelerator at Birmingham University, England. The results of this test indicated that relevant quantities of both bulk and sheet explosives can be detected against the background of diversely composed, highly cluttered cargo environments [9,10,11].

The issue of optimal detector type for a GRA-based EDS has long been the subject of intensive debate. Many claims and counter-claims have been made in favour of one or the other of the above-mentioned detector variants, but no systematic and comprehensive side-by-side evaluation has yet been presented. Bridging this information gap is the prime motivation for the present work and constitutes its principal subject.

## 2. THE GRA METHOD

### 2.1 Physical Principles of GRA

In GRA the nucleus $^{14}$N is raised to an excited state by absorbing an incident photon. The cross section for excitation of a nucleus to an excited state $E_R = 9.17$ MeV by absorption of a photon with energy E has a resonance behavior:

$$\sigma_{ABS}(E) = \pi\lambda^2 g \frac{\Gamma_\gamma \Gamma_T}{(E-E_R)^2 + \Gamma_T^2/4} \qquad \text{Eq.1}$$

where: $\Gamma_T$ is the total energy width of the level; the adopted average value = **128 eV** [12,13]

$\Gamma_\gamma$ is the partial width for the direct gamma ray transition to the ground state = **6.3 eV** [14]

$\lambda$ is the reduced wavelength corresponding to 9.17 MeV = **2.15x10$^{-12}$ cm**

g is a statistical spin factor: $g = \frac{2J_E + 1}{2(2J_G + 1)} =$ **5/6**

and $J_E = 2$, $J_G = 1$ are the total angular momenta of the excited and ground nuclear states, respectively

The resonant absorption cross-section (ignoring Doppler broadening) for $^{14}$N nuclei near the 9.17 MeV level is shown in Fig. 1, together with those for the more familiar non-resonant atomic attenuation processes (Compton and pair production).

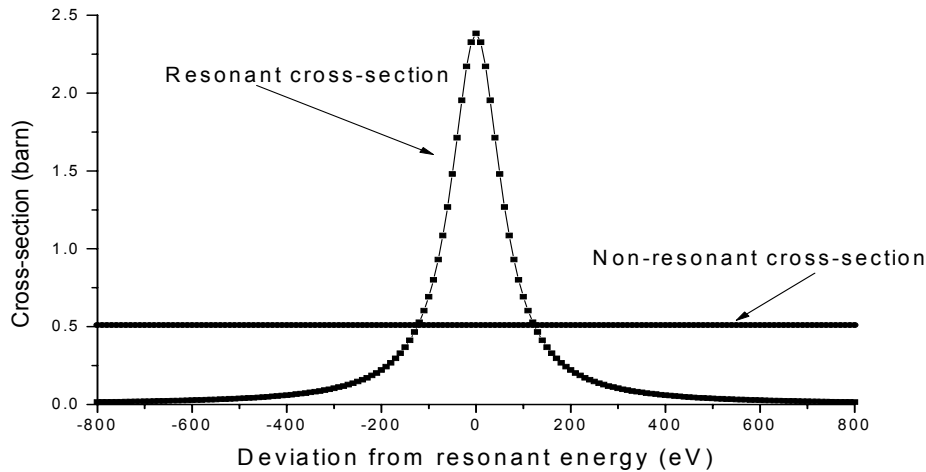

Fig. 1: Resonant absorption $\sigma_{ABS}(E)$ (square dots) and non-resonant (solid line) cross-sections for $^{14}$N at 9.17 MeV

In all materials the non-resonant attenuation varies very slowly with γ-ray energy (by only a few % over a range of ± 1 MeV around the resonant energy). At precisely the resonance energy = 9.176 MeV, the resonant absorption cross section is ~5 times higher than the non-resonant attenuation cross section in nitrogen, namely, **2.4 barns/atom**.

Specifically, at excitations in $^{14}$N above the proton separation energy of 7.55 MeV, the nucleus can decay either by emission of a gamma ray or by decomposition into $^{13}$C + proton. The cross sections for such reactions are, respectively:

$$\sigma_{\gamma\gamma} = \sigma_{ABS}(E)\frac{\Gamma_\gamma}{\Gamma_T} \quad \text{and} \quad \sigma_{\gamma p} = \sigma_{ABS}(E)\frac{\Gamma_p}{\Gamma_T} \quad \text{Eq.2}$$

where $\Gamma_p$ is the partial decay width for protons = **121.7 eV** [12)]

In GRA transmission imaging, the inspected object is scanned by a beam of γ-rays, a fraction of which represent on-resonance flux at 9.17 MeV, that is capable of being absorbed by $^{14}$N nuclei. Thus, in addition to the familiar non-resonant atomic processes (compton scattering and pair production) that attenuate the transmitted flux, such γ-rays will undergo an additional, nuclear resonance component of attenuation that is proportional to the line-integrated areal density of nitrogen in the line-of-sight from source to detector. By measuring the transmitted flux at energies on and off-resonance and normalizing appropriately, both the non-resonant (integral-absorber) and net resonant (total-nitrogen) components of attenuation can be extracted.

**2.2 Source of Resonant Radiation**

The ideal γ-ray source for nitrogen-GRA would be one which emits photons that are all concentrated in an energy interval of the order of the total level width around the 9.17 MeV $^{14}$N resonance, so that all of them undergo resonant attenuation if the inspected object contains a nitrogenous material.

**2.2.1 The $^{13}$C(p, γ)$^{14}$N reaction**

It turns out that the γ-ray source closest to the ideal is the de-excitation spectrum of the excited $^{14}$N* 9.17 MeV level following proton capture via the reaction $^{13}$C(p,γ)$^{14}$N [15)]. This reaction is the inverse of the resonant absorption process described in section **2.1**. It occurs at a proton energy of **1.747 MeV** and the thick target reaction yield into 4π is calculated to be **~6x10$^{-9}$ γ/proton**. Since the lifetime of the 9.17 MeV level (**5.1x10$^{-18}$ s**) is very short compared to ion stopping times (typically ~1x10$^{-12}$ s) the emission of the gamma-ray occurs during the recoil of the excited $^{14}$N nucleus, resulting in Doppler-shifting of the γ-ray. At the resonant angle **$\theta_R$ = 80.66°** with respect to the proton beam, the nuclear-recoil energy losses that occur during emission and absorption of the γ-ray by the $^{14}$N nucleus are precisely compensated by the Doppler-shifted energy component.

**2.2.2 The 9.17 MeV emission line and the spectrum incident on inspected objects and detectors**

The resonant photons are emitted with axial symmetry relative to the proton beam, forming an angular cone of width Δθ (FWHM) centered around the resonant angle $\theta_R$. Due to the Doppler-shift, Δθ is determined by a convolution of the intrinsic level width $\Delta\theta_{nucl}$ (~0.18°) with beam and target-related broadening effects. It is precisely these broadening factors that determine the suitability of a particular radiation source and detector for GRA applications. The value of Δθ has been determined by several investigators [12,14)] and also by the SNRC/LANL collaboration [13)]. All experiments consistently show that, despite employing beams of excellent quality (with respect to proton beam divergence and energy resolution) and viewing the emission line through a narrow collimator slit of 0.25°, the angular aperture size required to include most of the resonant photons is ~**0.75°**. This aperture corresponds to an energy spread of **520 eV**, since the Doppler variation of the γ-energy is calculated from capture reaction kinematics to be **694 eV/degree** at $\theta_R$. Such an energy spread is a factor of ~4 larger than the total intrinsic width of the level $\Gamma_T$.

We shall now examine the sources of broadening of the 9.17 MeV emission line, for which the known contributions are:

- Nuclear level width - **128 eV**

- Proton beam energy resolution (multiple scattering) - **few eV broadening per keV**

- Proton beam optics (parallelism, spot size) - ~**120 eV**

- Doppler broadening by vibrating target nuclei at elevated target temperatures - **40 eV (300°K), 80 eV (1300°K)**

Clearly, none of the above broadening factors can account for the observed 520 eV width of the emission line. Moreover, such broadening effects are typical of resonance absorption experiments populated in the (p,γ) reaction, as observed on other levels in light nuclei [16,17]. In all these cases, the broadening presumably results from atomic excitation processes in the target atom that are concomitant with proton capture by its nucleus. Such combined atomic/nuclear excitations have indeed been observed in (p,γ) reactions using proton beams with ultra-high resolution of ~15 eV [18].

In order to determine the incident 9.17 MeV spectrum on the inspected objects and detectors, the broadened emission line f(E) must be convoluted with an energy shift function g(u) representing the angular aperture, as follows:

$$\Phi(E) = k \int_{-\Delta E/2}^{\Delta E/2} f(E-u)\, g(u)\, du \qquad \text{Eq. 3}$$

where, in the present case, Φ(E) is the said incident 9.17 MeV spectrum, ΔE is the Doppler energy shift corresponding to an angular aperture of 0.75° and k is a proportionality constant. It is assumed that g(u) is constant within ΔE and that f(E) is no longer lorenzian but has a gaussian form with a FWHM of 520 eV.

The shape of the incident 9.17 MeV spectrum, calculated with the above procedure, is shown in Fig. 2, along with the shape of the absorption cross-section. As can be observed, the former is much broader than the latter. Consequently, only **~25%** of the incident photons are useful for resonant absorption.

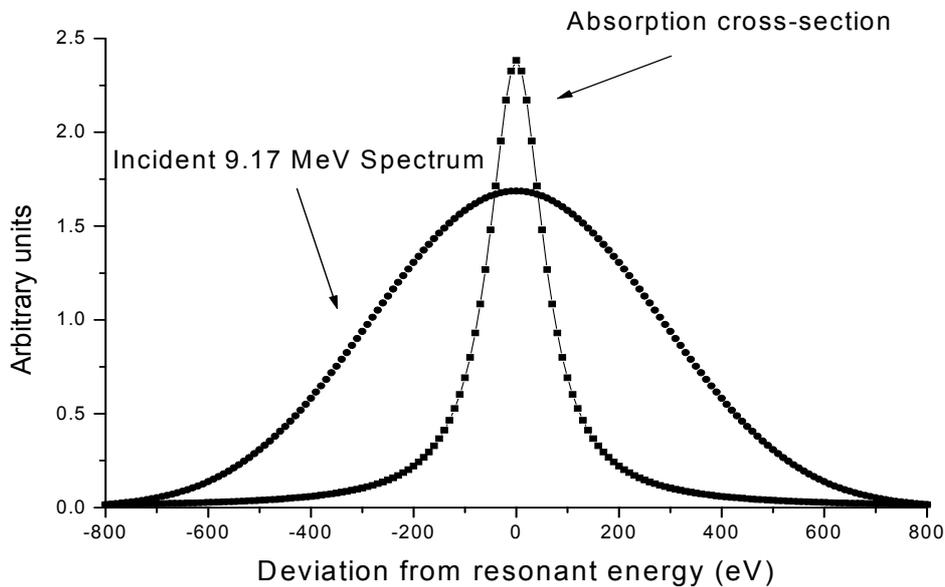

Fig. 2: 9.17 MeV incident spectrum Φ(E) (circles) and the resonant absorption cross-section (squares)

### 2.2.3 Normalization procedures and $^{13}$C targets
In order to factor out the non-resonant attenuation for the 9.17 MeV photons, it is necessary to measure it separately from the total (resonant + non-resonant) attenuation. To date, two variants have been employed for this purpose:

- Measurement of the attenuation at an off-resonant angle
- Introduction of an additional γ-ray line at a close non-resonant energy, that undergoes non-resonant attenuation only

The first approach is the classical one and has been used in most resonant absorption experiments. However, it involves two separate measurements: at angles on and off resonance. The TRIUMF/Grumman group [7,8] developed a method in which the two measurements are performed simultaneously. This is done using a position-sensitive detector that covers both the resonant and a neighbouring non-resonant angular range. Thus, one section of the detector measures the total (resonant + non-resonant) attenuation and another - the non-resonant attenuation. The principal advantage of this method lies in employing a simple thin $^{13}$C target bombarded by protons at the capture resonance that generates just the 9.17 MeV γ-ray as radiation source. The principal disadvantage is that the two measurements are performed at different angles, so that the resonant and non-resonant photons sample adjacent but non-identical line integrals through the inspected object. Hence corrections for this must be applied in the image reconstruction procedure. Another disadvantage is that, in order to accommodate the on and off-resonance flux simultaneously, the angular aperture of the fan beam needs to be wide, with the result of γ-rays being scattered by the inspected object from the non-resonant region to the resonant one and vice versa.

The second approach, employed by LANL and Soreq, is based on a simultaneous measurement of the 9.17 MeV γ-ray together with supplementary radiation emitted from the target at a different energy. The principal advantage of this method is that the two radiations view the inspected object in exactly the same geometry. The principal disadvantage is that, if the difference in energies is large, the factoring out of the non-resonant attenuation component is not accurate for all materials (notably for high-Z absorbers). Moreover, the target is more complex and the detector needs to be capable of distinguishing between the two radiations.

The LANL group used a composite $^{13}$C/$^{19}$F target [6] which, in addition to the 9.17 MeV line, generates intense lines of 6.13, 6.92 and 7.12 MeV from the $^{19}$F(p,α)$^{16}$O reaction. A typical target consisted of a 150 μg/cm$^2$ thick $^{13}$C layer and a 65 μg/cm$^2$ thick layer of barium fluoride, separated by a thin (40 μg/cm$^2$) layer of hafnium. The large energy difference between the 9.17 MeV and the $^{16}$O lines can prove problematic for factoring out the non-resonant attenuation quantitatively, as noted above.

The Soreq group used a thick (50 μm) $^{13}$C target [19,20] as a source of resonant and non-resonant radiation. Such a target exhibits the integral yield of all nuclear reactions of p + $^{13}$C from the initial bombarding energy down to zero. The γ-rays emanate exclusively from resonances in the proton capture reaction at different beam energies. Thus, in addition to the 9.17 MeV radiation, this target generates a γ-ray at 8.06 MeV and a continuum of γ-rays from 7.5 to 9 MeV that represent a series of weaker resonances. Neither the 8.06 MeV line nor the continuum radiation exhibit resonant interactions with nitrogen. This approach has the advantage of producing γ-rays closer to 9.17 MeV than those employed by LANL, but, since the $^{13}$C layer is pyrolytically-deposited on a natural graphite backing, power dissipation, radiation damage and target stability issues will require further study for eventual field systems.

## 2.3 Detectors
In this section, we present the geometrical layout, mode of operation and calculated efficiencies for the three detector variants. The efficiencies are consistently quoted for the angular aperture of 0.75°.

### 2.3.1 Non-resonant detectors
As mentioned above, the non-resonant detectors hitherto employed in GRA were BGO detectors developed by the LANL and TRIUMF/ Grumman groups, each with a distinct configuration and related response characteristics.

The array of detectors in the LANL setup [6] consisted of 64 BGO detectors, each coupled to a photomultiplier (PM) tube. The shape of the detector was a parallelepiped of dimensions 50 x 29 x 90 mm$^3$ (height x width x length). A 20 mm wide collimating slit is positioned in front of the array to form an aperture that corresponds to the 0.75° angular width of the resonant photon fan beam. The 50 x 29 mm$^2$ detector face is presented normal to the γ-ray beam, with the long (90 mm) dimension in the efficiency direction. Pixel size is thus defined by the detector dimensions and the slit aperture to be 20 x 29 mm$^2$. To assure stable operation of the BGO detectors a constant temperature is required. This was achieved by cooling the array with water at 17°C. The quoted average energy resolution was **15%**. Such resolution is sufficient for adequate separation of the 9.17 MeV line from the 6.13, 6.92 and 7.12 MeV lines used for normalization. Since the intrinsic detection efficiency at 9.17 MeV for the above detector configuration is not quoted, we have performed a Monte Carlo (MCNP) calculation of the spectral response of the detector [21]. The total intrinsic detector efficiency and the peak-to-total (full, single and double escape peaks) ratio were found to be 93% and 59% respectively, resulting in a 9.17 MeV photon detection efficiency of **55%.**

The TRIUMF/Grumman detector [22] is made of a segmented block of BGO crystal coupled to 4 PM tubes. In this fashion position-sensitivity within the block is achieved. The dimensions of the block are 50 x 50 x 50 mm$^3$. The detector is segmented by cutting the block into 12 x 12 subvolumes. This results in crystal elements approximately 4 x 4 x 50 mm$^3$ in size. The segmented block is viewed with four PM tubes, which determine the position of interaction. The energy spectrum is determined by summing the light from the four tubes. The quoted interaction probability, or the total intrinsic detector efficiency for 9.17 MeV photons was **77%** [22], however, again, no peak efficiency is quoted. Our MCNP calculation for a BGO block of the above dimensions yielded a total intrinsic efficiency of 77% and peak (full, single and double escape) efficiency ranging fom 50% in the central segment to 16% in the corner segment. Inevitably, the experimental peak efficiency is much worse, reflecting severe degradation of the energy resolution due to poor light collection in the long, narrow crystals and variation of the collected light fraction with depth of interaction. As shown by the authors, the energy spectrum is indeed strongly dependent on the position of interaction, the maximal spectrum degradation occurring in the corner segment. In theory, one could collect all the events in the spectrum, that are due to the 9.17 MeV interaction in the crystal. In practice, however, one has to define an event discrimination threshold as high as ~8 MeV, in order to prevent collection of scattered events within the larger γ-beam aperture employed here. We estimate that the upper limit for detection efficiency (averaged over all segments) achievable here is **~30%**.

### 2.3.2 Resonant Detector

In order to single out the resonant photons from the total radiation transmitted through the inspected object, SNRC has developed resonant-response detectors that are selectively sensitive to photons in the 9.17 MeV ± 130 eV interval. The detectors developed here consist of nitrogen-rich liquid scintillators (NRLS) [3,4,5], in which the incident photons react resonantly with nitrogen via the resonant photonuclear (γ,p) reaction, the inverse of the proton capture (p,γ) reaction on $^{13}$C. The resulting, internally-produced 1.5 MeV proton events are detected with 100% efficiency and can be distinguished via pulse shape discrimination (PSD) from the numerous electrons and positrons (leptons) produced by γ-rays that do not interact resonantly. PSD is based on the difference in scintillation decay times from high ionization (proton) and low ionization (lepton) events, which can be translated by an appropriate analog electronic circuit to time peaks that are separated by ~15 ns. Counting the proton signal provides a measure of the total transmission attenuation (resonant and non-resonant). The non-resonant attenuation is determined in a simultaneous measurement, by counting the lepton signal.

The NRLS has a density of **$\rho_T$ = 0.9 g/cm$^3$** and its nitrogen content is **30%** by weight, so that the nitrogen density is **$\rho_N$ = 0.27 g/cm$^3$**. The detector consists of a pyrex vessel of parallelepiped shape, filled with the liquid, 20 x 20 x 240 mm$^3$ in dimensions. The vessel is coupled via a parabolic perspex light-guide to a 50.8 mm PM tube and the entire assembly is enclosed by a light-tight cover. The 20 x 20 mm$^2$ detector face is presented normal to the γ-ray beam, forming a pixel of these dimensions, with the long (240 mm) dimension in the efficiency direction. The detector array for the 1998 LD-3 container test consisted of 61 such detectors.

Fig. 3 shows a two-dimensional (scintillation decay-time vs. pulse-height) spectrum obtained using the resonant detector irradiated by 9.17 MeV photons on and off the resonant angle.

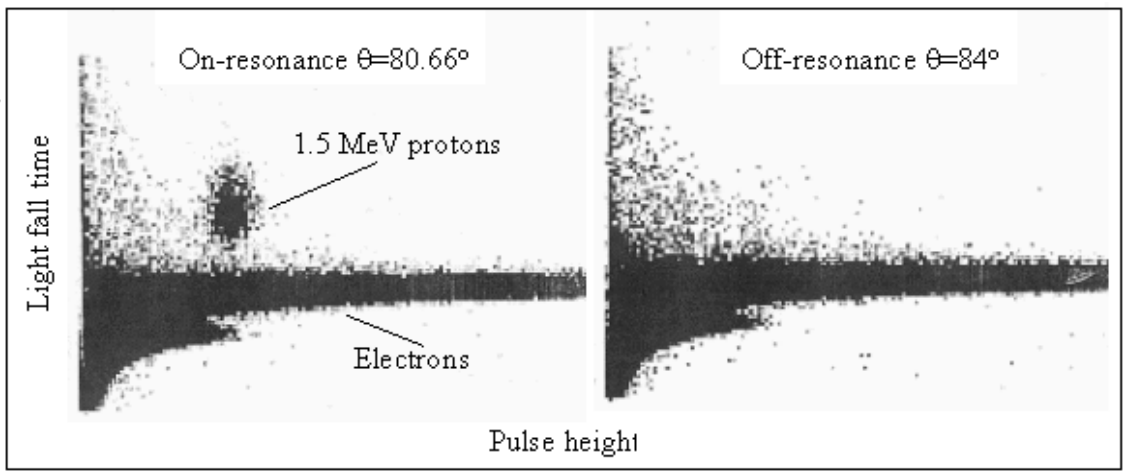

Fig. 3: Two-dimensional display of the resonant NRLS detector response to 9.17 MeV γ-rays, on and off the resonant angle.

One can observe from Fig. 3 that the 1.5 MeV proton events created in the $^{14}$N(γ,p)$^{13}$C reaction by resonant photons are very well separated from the continuum of leptons created by all photons in the incident spectrum. By electronically selecting a two-dimensional region of interest around the proton peak one can count the proton events only, with a signal- to-background ratio (for proton to lepton events) exceeding 15:1. The lepton events which determine the non-resonant attenuation are counted in a separate channel above a 3.5 Mev pulse-height threshold.

The response of the resonant detector to photons is energy-dependent and its resonant efficiency is represented by:

$$\text{eff}_R(E) = \frac{\mu_{\gamma p}(E)\rho_N}{\Sigma_T(E)}\left[1 - \exp(-\Sigma_T(E)d)\right] \qquad \text{Eq. 4}$$

where $\mu_{\gamma p}(E)$ is the mass attenuation cross section for the (γ,p) reaction in the detector, $\Sigma_T(E) = \mu_{\gamma p}(E)\rho_N + \mu_{NR}\rho_T$ is the total macroscopic cross-section for interaction of the 9.17 MeV photons in the detector medium, $\mu_{NR}$ is the conventional mass attenuation coefficient and d is the length of the detector (24 cm).

Fig. 4 shows the detection efficiency of the NRLS resonant detector as function of γ-ray energy. As can be observed, the efficiency is relatively high (**40%**) at resonance and decreases rapidly as the energy deviates from the resonant energy. The efficiencies of the LANL and TRIUMF/Grumann BGO detectors are shown here for comparison.

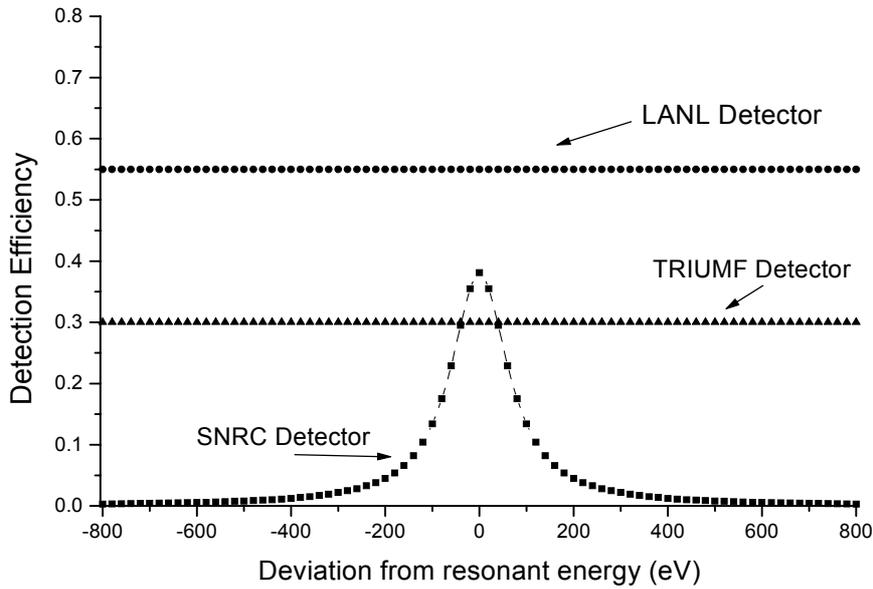

Fig. 4: Intrinsic detection efficiency for LANL BGO, TRIUMF/Grumman BGO and SNRC NRLS detectors

## 3. COMPARISON OF DETECTORS

In this section, we present intrinsic figures-of-merit (*F.O.M's*) for evaluating and comparing the single-pixel performance of the three detector variants. These *F.O.M's* are consistently quoted for the relevant angular aperture of 0.75°.

Specifically, when comparing relative merits of GRA detectors with resonant and non-resonant response, it is important to take into consideration how each of these detectors responds to the incident radiation, how well it singles out the resonant photon events that carry the nitrogen attenuation information and rejects non-resonant photon events that carry only the atomic attenuation information. Once this has been clarified, it is relatively easy to define a first-order *F.O.M*. By first order, we refer to the approximation that there are no artifacts of any of the following types:

- Energy dependent normalization
- Spatial sampling
- Image noise of any kind except Poisson counting statistics
- Non-resonant forward-scattered radiation in the transmission image

In this approximation, the *F.O.M* is defined in Eq. 5 below. It represents a single-pixel measure for statistical significance of GRA data and thus has a direct bearing on the throughput, probability of detection and false alarm rate achievable with the detector.

$$F.O.M = \text{Contrast\_sensitivity} \cdot \sqrt{\text{Number\_of\_detected\_events}} \qquad \text{Eq. 5}$$

By Contrast_sensitivity is meant the measured net resonant attenuation per g/cm$^2$ of nitrogen traversed in the inspected object (essentially, the mean resonant attenuation coefficient). The Number_of_detected_events is defined as number of detected 9.17 MeV events per given incident photon intensity Φ(E). From Eq. 5 it is clear that the Contrast_sensitivity has a greater effect on the *F.O.M* than the Number_of_detected_events.

Table I below presents the analytical expressions for the above parameters. The Contrast_sensitivity expressions are given here for a thin nitrogenous absorber, i.e., taking a linearized approximation of the exponential resonant attenuation. Such an approximation is reasonable since, in most operational scenarios, the nitrogen thickness traversed does not exceed ~3g/cm$^2$. $\mu_A(E)$ is the mass resonant attenuation of nitrogen defined as: $\mu_A(E) = \sigma_{ABS}(E) \times N_{AVOGADRO}/14$. All other values have been defined previously.

Table 1. Expressions for Contrast_sensitivity and Number_of_detected_events

| Parameter | Resonant Detector | Non-Resonant Detector |
|---|---|---|
| Contrast_sensitivity (thin absorber) | $\dfrac{\int_{-\infty}^{\infty} \mu_A(E)\, \Phi(E)\, \text{eff}_R(E)\, dE}{\int_{-\infty}^{\infty} \Phi(E)\, \text{eff}_R(E)\, dE}$ | $\dfrac{\int_{-\infty}^{\infty} \mu_A(E)\, \Phi(E)\, dE}{\int_{-\infty}^{\infty} \Phi(E)\, dE}$ |
| Number_of_detected_events | $\int_{-\infty}^{\infty} \Phi(E)\, \text{eff}_R(E)\, dE$ | $\text{eff}_{NR} \int_{-\infty}^{\infty} \Phi(E)\, dE$ |

As can be observed in Table 1, the Contrast_sensitivity for the non-resonant detector is independent of detector parameters and depends only on the spectrum of the incident radiation. We have calculated the Contrast_sensitivity for both types of detectors by evaluating the integrals numerically, using Φ(E) determined in section 2.2.2, i.e., corresponding to an intrinsic emission line width of 520 eV and a collimator aperture of 0.75°. The resulting calculated Contrast_sensitivities were **0.025** and **0.058** cm$^2$/g for the LANL non-resonant and SNRC NRLS resonant detectors, respectively. These agree with the experimental values of **0.026** [23] and **0.054** [3] respectively, obtained with electrostatic accelerators for similar collimator apertures.

In general, it is to be noted that any degradation in target or accelerator beam quality will have more severe consequences for the *F.O.M's* of non-resonant detectors than for those of resonant detectors. Thus, for the resonant detector, the Contrast_sensitivity is invariant to such changes, whereas the Number_of_detected_events will decrease in proportion to the emission-line broadening. In comparison, for the non-resonant detector, the Contrast_sensitivity will decrease in proportion to the beam and target related emission-line broadening, whereas the Number_of_detected_events will remain invariant. Indeed, the experimental evidence is that the Contrast_sensitivity of the LANL BGO detector obtained with an RFQ accelerator beam of ~20 keV FWHM energy resolution was reduced from **0.026** to **0.015** cm$^2$/g [23].

For a quantitative evaluation of the influence of the accelerator/target-induced incident line broadening we have calculated the Contrast_sensitivity for non-resonant and resonant detectors, treating the intrinsic emission linewidth (f(E) of Eq.3) FWHM as a parameter. Fig. 5 shows the Contrast_sensitivity for resonant and non-resonant detectors vs. the FWHM of the intrinsic emission line. In all cases the collimator aperture was maintained at 0.75°.

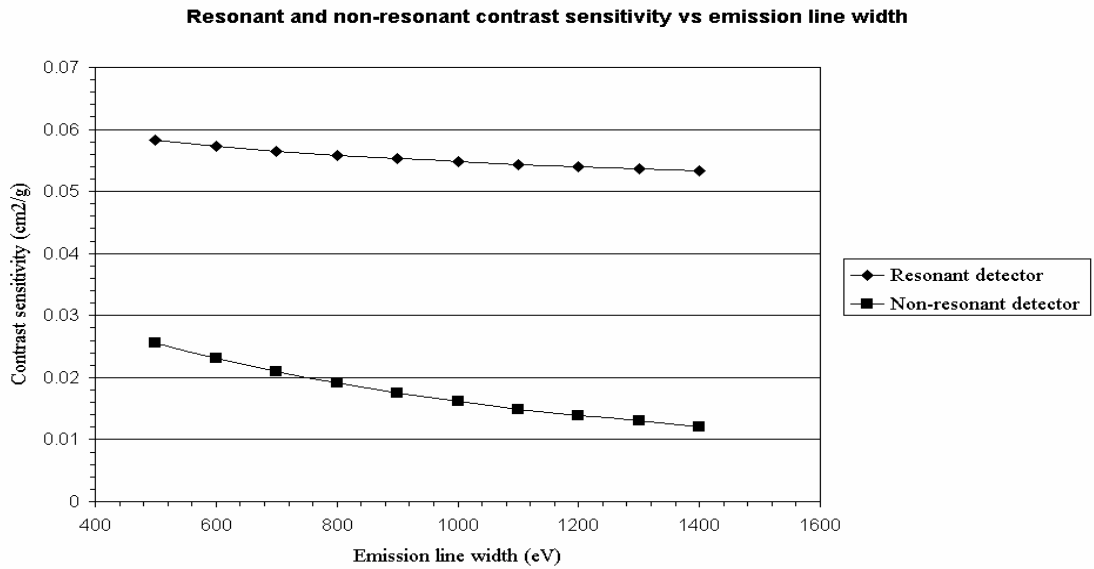

Fig. 5: Non-resonant and resonant detector Contrast_sensitivity vs the FWHM width of the intrinsic emission line

Fig. 5 clearly shows that the non-resonant detector is inherently more sensitive to changes in the width of the emission line than the resonant detector. For example, if the emission line is broadened to 1100 eV, the resonant detector Contrast_sensitivity becomes 3.6 times higher than that of its non-resonant counterpart.

Obviously, the next step is to perform the comparison between non-resonant and resonant detectors at the level of the *F.O.M* itself. To this end, we have calculated the *F.O.M* for each detector vs the angular aperture of the collimator. As the aperture size varies, the two types of detector will respond differently. Fig 6 shows the *F.O.M* ratios for an NRLS resonant detector to a non-resonant detector of LANL and TRIUMF/Grumman design, as function of collimator aperture.

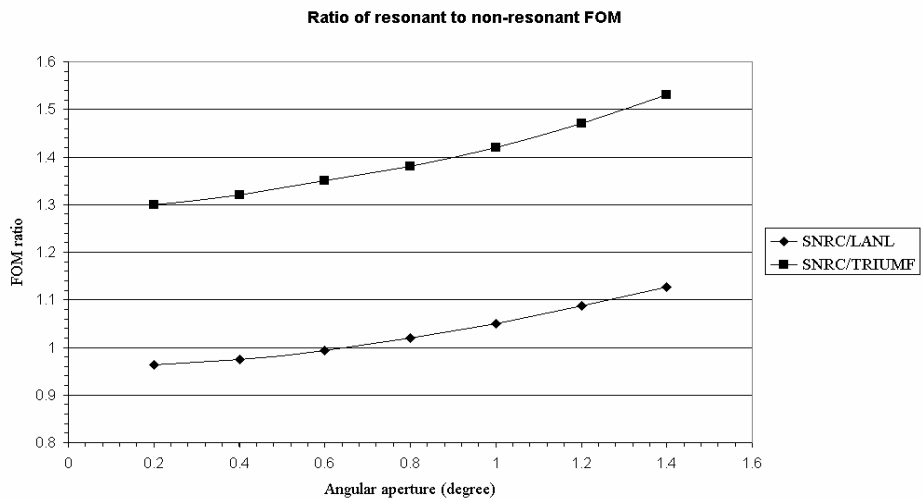

Fig. 6: *F.O.M* ratios of the SNRC resonant detector to the LANL (diamonds) and TRIUMF/Grumman (squares) detectors

Fig. 6 shows that the ratio of single-pixel *F.O.M* for the SNRC NRLS resonant and LANL-BGO non-resonant detectors is approximately unity at an angular aperture of 0.75°, indicating that, from a purely statistical point of view, the performance of these two detectors is comparable. In contrast, the performance of the resonant detector is clearly superior to that of the TRIUMF/Grumman non-resonant detector at all aperture sizes.

## 4. CONCLUSIONS

A clear-cut conclusion is that resonant-response detectors invariably exhibit higher Contrast_sensitivity than their non-resonant counterparts. Typically, the latter is at least a factor of ~2.5 higher for NRLS than for BGO. This feature of NRLS unequivocally renders it the detector of choice when small quantities of explosives need to be detected. This type of detector is also free of normalization and target stability problems. Its main shortcoming is relatively low efficiency, since it has hitherto proved difficult to load more than ~30% nitrogen (weight per weight) into the active medium without impairing PSD performance by scintillation quenching.

Another conclusion is that, from a purely statistical point of view, the performances of NRLS and LANL-BGO detectors are comparable. However, this is valid only to the extent that the GRA data and images do not exhibit noise or artifacts of any kind except Poisson counting statistics. Not surprisingly, under real measuring conditions, one does expect (and indeed experimentally observes [22]) additional sources of data and image noise, that are not entirely reflected in the single-pixel *F.O.M's*. Prominent among these are:

- degradation or instability of accelerator beam quality (primarily with respect to beam energy resolution, spot size and optics at target)
- target deterioration under protracted beam bombardment
- low-angle γ-ray scattering (highly significant when scanning massive, low-Z objects)

The first two of these artifacts clearly degrade BGO data more severely than NRLS data. This is inherent in the resonant vs. non-resonant response. With respect to the third, there is no marked preference for one detector type over another, except that the TRIUMF/Grumman BGO variant appears somewhat more vulnerable than the other two. Essentially, the problem needs to be tackled by ensuring stringent pre-collimation of the γ-ray fan-beam.

A further conclusion is that NRLS is superior to the TRIUMF/Grumman detector at any aperture size (see Fig. 6). The dominant factor here is the relatively low photopeak efficiency that characterizes a segmented BGO detector. Another problem that is particularly severe with this detector is the influence of the low angle γ-ray scattering by the object from the non-resonant angle region to the resonant angle, which further reduces the resonant contrast.

In summary, we conclude that resonant detectors provide a considerably greater margin of sensitivity. They are also less dependent on accelerator beam quality and their lower detection efficiency is offset by the higher Contrast_sensitivity. Moreover, although not yet commercially available, NRLS detectors are markedly cheaper than BGO detectors.

Finally, it should be borne in mind that the performance of an eventual field EDS based on GRA will also reflect factors that cannot entirely be incorporated within the single-pixel *F.O.M* developed and discussed here. These pertain primarily to issues of image quality and the specific mode of inspection implemented (single-view or few-view radiography, few-view or multi-view tomography). They will be dealt with in a separate publication.